\documentclass[twoside,fleqn]{article}
\usepackage{espcrc2}
\usepackage[dvips]{epsfig}

\title{
  \vskip-2.0ex\hbox to 6.25in {{\normalsize \hfil FERMILAB-Conf-98/340-T}}
  $B\rightarrow Dl\nu$ form factors and the 
  determination of $|V_{cb}|$}

\author{
  S.~Hashimoto\address{Fermilab, P.O.~Box 500, Batavia, IL 60510},
  A.X.~El-Khadra\address{Physics Department, University of
    Illinois, Urbana, IL 61801},
  A.S.~Kronfeld$^{\rm a}$, P.B.~Mackenzie$^{\rm a}$,
  S.M.~Ryan$^{\rm a}$ and J.N.~Simone$^{\rm a}$
  }

\begin{document}

\begin{abstract}
  The zero recoil limit of the $B\rightarrow D l\nu$ form
  factors is calculated on the lattice, which provides a
  model-independent determination of $|V_{cb}|$.
  Considering a ratio of form factors, in which the bulk of
  statistical and systematic errors cancel, we obtain a
  precise result both for $h_+(1)$ and for $h_-(1)$.
\end{abstract} 

\maketitle

\section{Introduction}
\label{sec:Introduction}

For the determination of $|V_{cb}|$ through the exclusive
decay $B\rightarrow D^{(*)}l \nu$, the theoretical
calculation of the form factor 
$\mathcal{F}_{B\rightarrow D^{(*)}}(w)$, especially in the
zero recoil limit, is necessary.
Previously this has been done using the zero-recoil sum rule
\cite{Shifman_Uraltsev_Vainshtein_95} or the heavy quark
expansion of the form factor
\cite{Falk_Neubert,Ligeti_Nir_Neubert_94}. 
Both of these calculations, however, need to introduce
an assumption or a theoretical model to deal with 
hadronic effects away from the infinite quark mass limit.
In this talk, we present a lattice calculation of the form
factor $\mathcal{F}_{B\rightarrow D}(1)$, which can be used
for a model independent determination of $|V_{cb}|$,
including deviations from the heavy quark limit.

\section{$B\rightarrow Dl\nu$ Form Factors}
\label{sec:Form_Factors}

The differential decay rate of $B\rightarrow D l\nu$ is
proportional to the square of
\begin{equation}
  \label{eq:decay_rate}
  \mathcal{F}_{B\rightarrow D}(w) = h_+(w) -
  \frac{m_B-m_D}{m_B+m_D} h_-(w).
\end{equation}
$h_+(w)$ and $h_-(w)$ are form factors defined through
\begin{eqnarray}
  \label{eq:definition_of_the_form_factors}
  \lefteqn{\langle D(\mathbf{p'})| \mathcal{V}^{\mu} |
    \bar{B}(\mathbf{p})\rangle = \sqrt{m_B m_D} }
  \nonumber \\ 
  & & \times
  [h_+(w) (v+v')^{\mu} + h_-(w) (v-v')^{\mu}],
\end{eqnarray}
where $v_{\mu}=p_{\mu}/m_B$, $v_{\mu}'=p_{\mu}'/m_D$ and
$w=v\cdot v'$.
In the heavy quark mass limit, $h_{-}(w)$ vanishes, and
$h_{+}(w)$ agrees with the universal form factor $\xi(w)$
(the Isgur-Wise function), which is normalized in the zero
recoil limit $\xi(1)=1$ \cite{Isgur_Wise_89_90}. 

The $1/m_Q$ expansion may be used to describe the deviation
from the heavy quark limit.
At zero recoil, the expansion becomes 
\begin{eqnarray}
  \label{eq:1/m_Q-expansion_of_form_factors}
  h_+(1) & \!\! = \!\! & 1 - 
    c_+^{(2)} \left(\frac{1}{m_c}-\frac{1}{m_b}\right)^2 +
    O(1/m_Q^3), \\ 
  h_-(1) & \!\! = \!\! & 0 -
    c_-^{(1)} \left(\frac{1}{m_c}-\frac{1}{m_b}\right) +
    O(1/m_Q^2).
\end{eqnarray}
The absence of the $O(1/m_Q)$ term in the expansion of
$h_+(1)$, which is a part of the Luke's theorem
\cite{Luke_90}, is a consequence of the symmetry under the
exchange of $m_c$ and $m_b$ (see
eq.(\ref{eq:definition_of_the_form_factors})), 
and this particular form of the expansion is also restricted
by the symmetry (anti-symmetry for $h_-(1)$).
Our task is to determine the coefficients $c_+^{(2)}$ and
$c_-^{(1)}$, for which there has been no model independent
calculation.

\section{$h_+(1)$}
\label{sec:h_+}

The vector current $\mathcal{V}_{\mu}=\bar{c}\gamma_{\mu}b$
appearing in eq.(\ref{eq:definition_of_the_form_factors})
must be related to the lattice counterpart $V_{\mu}^{latt}$
using the perturbative relation
$\mathcal{V}_{\mu}=Z_V V_{\mu}^{latt}$. 
This is true even for the equal mass ($m_c=m_b$) case,
because the lattice (local) vector current is not
conserved. 
Without  a two-loop calculation, this perturbative matching
could in principle
be a source of large systematic uncertainty of
$O(\alpha_s^2)\sim 5\%$, which is too large to obtain 
the precision we seek
 for the form factor ($<5\%$).
The statistical error in the lattice calculation would also be
a problem, if we employed the usual method to extract the
matrix element from three-point correlator.

To reduce these errors, we define a ratio at zero recoil
\begin{eqnarray}
  \label{eq:R^B->D}
  R^{B\rightarrow D} & \!\! = \!\! & 
  \left[
  \frac{
    \langle D|\mathcal{V}_0|B\rangle 
    \langle B|\mathcal{V}_0|D\rangle }{
    \langle D|\mathcal{V}_0|D\rangle
    \langle B|\mathcal{V}_0|B\rangle }
  \right]^{cont}
  \nonumber \\
  & \!\! = \!\! &
  \frac{h_+^{B\rightarrow D}(1) h_+^{D\rightarrow B}(1)}{
        h_+^{D\rightarrow D}(1) h_+^{B\rightarrow B}(1)}
  = |h_+^{B\rightarrow D}(1)|^2,
\end{eqnarray}
where we used the property
$h_+^{D\rightarrow D}(1)=h_+^{B\rightarrow B}(1)=1$ derived
from current conservation.
$R^{B\rightarrow D}$ may be related to the lattice
counterpart
\begin{equation}
  \label{eq:R^B->D_latt}
  \frac{Z_{V^{cb}} Z_{V^{bc}} }{
        Z_{V^{cc}} Z_{V^{bb}} } \times
  \left[ \frac{
    \langle D|V_0|B\rangle \langle B|V_0|D\rangle }{
    \langle D|V_0|D\rangle \langle B|V_0|B\rangle }
  \right]^{latt}.
\end{equation}
The ratio of matching factor can be safely evaluated with
perturbation theory, since a large cancellation of
perturbative coefficients takes place in the ratio.
The one-loop calculation is discussed in a separate talk
\cite{Kronfeld_Hashimoto_98}.

To calculate the hadronic amplitude,
we define
\begin{equation}
  \label{eq:R^B->D(t)}
  R^{B\rightarrow D}(t) =
  \frac{C^{DV_0B}(t) C^{BV_0D}(t)}{C^{DV_0D}(t) C^{BV_0B}(t)}
  \rightarrow R^{B\rightarrow D},
\end{equation}
where $C^{DV_0B}(t)$ is a three-point correlator, whose
initial and final states are fixed at $t=0$ and $t=N_t/2$
respectively and the vector current is moved to find a
plateau.
Our calculation is done on a 12$^3\times$24 lattice at
$\beta=5.7$. 
The Fermilab action \cite{El-Khadra_Kronfeld_Mackenzie_97}
is used for heavy quark with $c_{sw}=1/u_0^3$.

\begin{figure}[t]
  \begin{center}
    \vspace*{-5mm}
    \epsfxsize=62mm \epsfbox{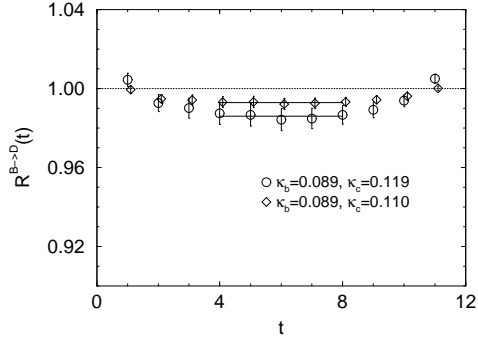}
    \vspace*{-9mm}
    \caption{$R^{B\rightarrow D}(t)$ as a function of $t$.}
    \label{fig:R^B->D}
  \end{center}
  \vspace*{-10mm}
\end{figure}

Figure \ref{fig:R^B->D} shows a nice plateau for the
ratio $R^{B\rightarrow D}(t)$.
Even with only 200 configurations the statistical error is
remarkably small ($< 1\%$), 
because of the cancellation of statistical fluctuations in the
ratio. 

\begin{figure}[t]
  \begin{center}
    \vspace*{-5mm}
    \epsfxsize=62mm \epsfbox{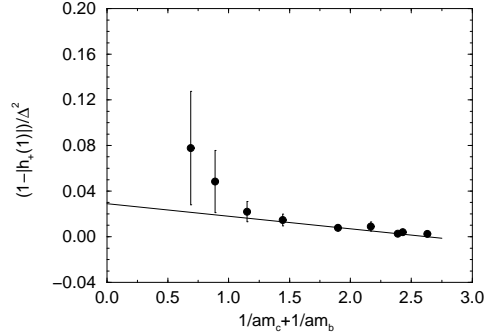}
    \vspace*{-9mm}
    \caption{
      $(1-|h_+(1)|)/(1/m_c-1/m_b)^2$ versus $1/am_c+1/am_b$.}
    \label{fig:h+_heavy_mass_coeff}
  \end{center}
  \vspace*{-10mm}
\end{figure}

Fitting the plot with a constant we obtain $|h_+(1)|^2$ for
each combination of initial and final heavy quark masses.
To fix the parameter in the $1/m_Q$ expansion, we choose six
values of the heavy quark mass covering the physical $m_b$
and $m_c$, and fit the results with the form
\begin{eqnarray}
  \label{eq:1/m_Q-expansion_of_h+}
  \lefteqn{ h_+(1) = 1 - 
    c_+^{(2)} \left(\frac{1}{m_c}-\frac{1}{m_b}\right)^2 }
  \nonumber \\
   & & +
   c_+^{(3)} \left(\frac{1}{m_c}+\frac{1}{m_b}\right) 
             \left(\frac{1}{m_c}-\frac{1}{m_b}\right)^2,
\end{eqnarray}
where  the $O(1/m_Q^3)$ term is required to explain the data.
Figure \ref{fig:h+_heavy_mass_coeff} shows the quantity
$(1-h_+(1))/(1/m_c-1/m_b)^2 =
c_+^{(2)}-c_+^{(3)}(1/m_c+1/m_b)$, from which we extract the
coefficients $c_+^{(2)}$ and $c_+^{(3)}$.
Our result is $c_+^{(2)}=0.029(11)$ and $c_+^{(3)}=0.011(4)$.
In physical units we obtain
$c_+^{(2)}=(0.20(4)\mbox{GeV})^2$ and
$c_+^{(3)}=(0.26(3)\mbox{GeV})^3$.
Shifman \textit{et al.}
\cite{Shifman_Uraltsev_Vainshtein_95} derived a bound 
$c_+^{(2)} > (\mu_{\pi}^2-\mu_G^2)/2 
 = (0.26^{+0.09}_{-0.12} \mbox{GeV})^2$ 
using the zero-recoil sum rule. \footnote{For the
   hadronic parameters we used $\mu_{\pi}^2=0.5(1)
   \mbox{GeV}^2$ and $\mu_G^2=0.36\mbox{GeV}^2$.}
Our result is consistent with this bound within errors.

\section{$h_-(1)$}
\label{sec:h_-}

$h_-(1)$ cannot be obtained from the matrix element at zero
recoil. 
We introduce a finite (but small) $D$ meson momentum
$\mathbf{p'}$ and define a ratio
\begin{eqnarray}
  \label{eq:R_Vi/V0}
  \lefteqn{
    R^{B\rightarrow D}_{V_i/V_0} = 
    \frac{\langle D(\mathbf{p'})|V_i|B(\mathbf{0})\rangle}{
          \langle D(\mathbf{p'})|V_0|B(\mathbf{0})\rangle}
  } \nonumber \\
  & = & \frac{1}{2} v'_i
  \left( 1 - \frac{h_-(w)}{h_+(w)} \right)
  \nonumber \\
  &   & \times
  \left[ 1 - \frac{1}{4}
             \left( 1 - \frac{h_-(w)}{h_+(w)}
             \right) \mathbf{v'}^2 + \cdots
  \right],
\end{eqnarray}
where we use the definition of the form factors and expand
in small $\mathbf{v'}^2$.
It is more convenient to define a double ratio
\begin{eqnarray}
  \label{eq:R_Vi/V0^(B->D)/(D->D)}
  \lefteqn{
    R^{(B\rightarrow D)/(D\rightarrow D)}_{V_i/V_0} = 
    R^{B\rightarrow D}_{V_i/V_0} / R^{D\rightarrow D}_{V_i/V_0}
  } \nonumber \\
  & = &
  \left( 1 - \frac{h_-(w)}{h_+(w)} \right)
  \left[ 1 - \frac{1}{4} \frac{h_-(w)}{h_+(w)} \mathbf{v'}^2
    + \cdots \right].
\end{eqnarray}
The property of elastic scattering 
$h^{D\rightarrow D}_-(w)=0$ is used for the denominator.

\begin{figure}[t]
  \begin{center}
    \vspace*{-5mm}
    \epsfxsize=62mm \epsfbox{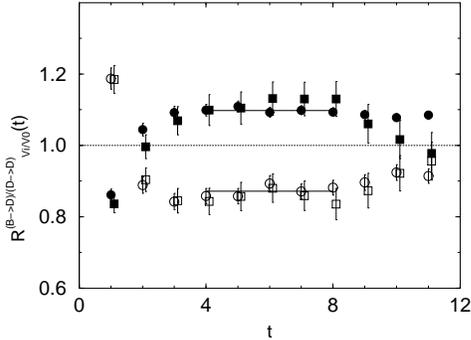}
    \vspace*{-9mm}
    \caption{
      $R^{(B\rightarrow D)/(D\rightarrow D)}_{V^i/V^0}(t)$ 
      for the final state momentum $(1,0,0)$ (circles) and
      $(2,0,0)$ (squares). 
      $B\rightarrow D$ (filled symbols) and $D\rightarrow B$
      (open symbols) are shown.
      }
    \label{fig:R_BDoverDD}
  \end{center}
  \vspace*{-10mm}
\end{figure}

Figure \ref{fig:R_BDoverDD} shows the corresponding ratio 
$R^{(B\rightarrow D)/(D\rightarrow D)}_{V_i/V_0}(t)$
on the lattice for two different values of the $D$ meson
momentum. 
As in the calculation of $h_+(1)$, the plateau is very clear
and we can extract $(1-h_-(w)/h_+(w))$ from this plot.
The small correction proportional to $\mathbf{v'}^2$ may be 
eliminated by extrapolating to the 
$\mathbf{v'}^2\rightarrow 0$ limit.
We also observe an antisymmetric property 
$h_-^{D\rightarrow B}(w) = - h_-^{B\rightarrow D}(w)$.

The heavy quark mass dependence can be obtained with a
similar strategy.
We fit our data with
\begin{eqnarray}
  \label{eq:1/m_Q-expansion_of_h-}
  \lefteqn{ h_-(1) = - 
    c_-^{(1)} \left(\frac{1}{m_c}-\frac{1}{m_b}\right) }
  \nonumber \\
   & & +
   c_-^{(2)} \left(\frac{1}{m_c}+\frac{1}{m_b}\right) 
             \left(\frac{1}{m_c}-\frac{1}{m_b}\right),
\end{eqnarray}
and obtain $c_-^{(1)}=0.23(3)$ and $c_-^{(2)}=0.06(1)$,
which correspond to $c_-^{(1)}=0.26(4)\mbox{GeV}$ and 
$c_-^{(2)}=(0.29(3)\mbox{GeV})^2$ in physical units.

\section{$\mathcal{F}_{B\rightarrow D}(1)$}
\label{sec:F}

The results with physical mass parameter are
\begin{eqnarray}
  \label{eq:results_h}
  h_+(1) & = &  1.016 \pm 0.003 \pm 0.002 \pm 0.006, \\
  h_-(1) & = & -0.112 \pm 0.014 \pm 0.011 \pm 0.025,
\end{eqnarray}
where $h_+(1)$ includes the one-loop correction +0.025(6)
\cite{Kronfeld_Hashimoto_98}.
Errors arise from statistics, mass parameter determination, 
and our estimate for
higher order perturbative correction in the order given.
Using (\ref{eq:decay_rate}) we obtain
\begin{equation}
  \label{eq:result_F}
  \mathcal{F}_{B\rightarrow D}(1) =
  1.069 \pm 0.008 \pm 0.002 \pm 0.025.
\end{equation}

\section{Conclusions}
\label{sec:Conclusion_and_Discussion}

Using a ratio in which a large cancellation of statistical
and systematic errors takes place, we have calculated the
$B\rightarrow D l\nu$ form factors very precisely, which
may lead to a better determination of $|V_{cb}|$.
In the error bar given above, however, we have not yet
included the discretization error and the effect of
quenching.  
It is our hope that the bulk of these errors also cancels in
the ratio. We leave these important issues for future study.

Our method can also be applied to the $B\rightarrow D^* l\nu$
form factor, which currently yields an experimentally more
precise determination of $|V_{cb}|$.

\end{document}